\documentclass[12pt]{article}
\usepackage[utf8]{inputenc}

\usepackage{amsmath}% Advanced Math Typesetting
\newtheorem{definition}{Definition}
\newtheorem{theorem}{Theorem}
\newtheorem{corollary}{Corollary}
\newtheorem{example}{Example}

\usepackage{amsfonts}% Unicode support
\usepackage{amssymb}
\usepackage{makeidx}
\usepackage{graphicx}% Add pictures to document
\usepackage{subcaption}
\usepackage{listings}% Source code formatting and highlighting

\usepackage{apacite}
\usepackage{natbib}

\usepackage{indentfirst}
\usepackage{caption}

\usepackage{ulem}
\usepackage{epsfig}

\usepackage{siunitx}
\usepackage{booktabs}
\newcolumntype{Y}{>{\centering\arraybackslash}X}

\usepackage{graphicx,psfrag,epsf}
\usepackage{enumerate}

\usepackage{url} % not crucial - just used below for the URL 
\usepackage{newtxmath}  
\usepackage{setspace}
\usepackage{bbm}
\usepackage{dsfont}
\usepackage{newtxtext}      
\usepackage{url}
\usepackage{mathtools}

\usepackage{authblk}
\usepackage[pdfborder={0 0 0}]{hyperref}% For email addresses

\title{Pay-As-You-Drive Insurance Pricing Model}
\author[a]{Safoora Zarei}
\author[b]{Ali R. Fallahi}
\begin{spacing}{.7}
\affil[a]{\small Shiraz Isalmic Azad University, Shiraz, Iran\\
\url{safooraeco@yahoo.com}}
\affil[b]{\small  Department of Mathematics and Computer Science, Amirkabir University, Tehran, Iran\\
\url{alirezafallahi@aut.ac.ir}}
\end{spacing}

\begin{document}
%*******************************************************************************************************
\thispagestyle{plain}

%\vspace{-8ex}}
\date{}

\maketitle
\begin{spacing}{1.00}
\noindent \textbf{Abstract}\ \ \ \ \ 
Every time drivers take to the road, and with each mile that they drive, exposes themselves and others to the risk of an accident. Insurance premiums are only weakly linked to mileage, however, and have lump-sum characteristics largely. The result is too much driving, and too many accidents. In this paper, we introduce some useful theoretical results for Pay-As-You-Drive in Automobile insurances. We consider a counting process and also find the distribution of discounted collective risk model when the counting process is non-homogeneous Poisson.
\\
\\
\noindent \textbf{Keywords}\ \ \ \ \   Cox Process, Martingales, Aggregate Risk Models, PAYD, Actuarial Mathematics.
\end{spacing}

\begin{spacing}{1.00}
\newpage

\section{Introduction}

In most developed countries, automobile insurance represents a considerable share of the yearly non-life premium collection. Because the majority of the incurred losses are usually very high. Therefore using this Insurance for car owners in some countries (such as Iran) is in force. Also, many attempts have been made in the actuarial literature to find a good model for calculating the premiums; for a review of the existing literature, we refer the interested readers, e.g., to \cite{Lemaire:1985}, \cite{Lemaire:1995}, or \cite{Frangos:2001}.

In a general case there are two different approaches:
\begin{itemize}
	\item Premium will be fixed for all of policyholders,
	\item Premium will be different for all of policyholders.
	
\end{itemize}

The first approach is unfair because, in this way, a policyholder who had an accident with a small size of loss (or did not have an accident) is being unfairly disadvantaged to a policyholder who had an accident with a big size of the loss. The second method is the right way and is the base of Bonus Malus Systems (BMS). BMS penalizes insurers responsible for one or more accidents by premium surcharges (or maluses), and rewarding claim-free policyholders by awarding them discounts (or bonuses). This is a very efficient way of classifying policyholders into cells according to their risk \citep[see][]{Denuit:2001}. A BMS calculates the premium applicable as a base premium, adjusted by a quantity (the bonus or malus) which depends on previous claims experience \citep[see][]{Taylor:1997}. There are three systems for adjusting the premiums:

\begin{itemize}
	\item BMS based on the frequency component,
	\item BMS based on the frequency and severity component,
	\item BMS based on the frequency component, severity component and individual characteristics	(such as sex, age, type of car, location and ...).
	
\end{itemize}

It is obvious that the third system is a generalization of two other systems \citep[see][]{Frangos:2001}. Moreover, \cite{Mahmoudvand:2009} have introduced a new generalization of BMS that covers all three types of BMS that mentioned above.

Recently, some researchers believe that the current lump-sum pricing of auto insurance is inefficient and inequitable \citep[see][]{Bordoff:2008}. Even according to the optimal BMS, drivers who are similar in claims and individual characteristics pay nearly the same premiums if they drive five thousand or fifty thousand miles a year. Considerable research indicates that annual crash rates and claim costs tend to increase with annual vehicle mileage \citep[see][]{Litman:2009}. This pricing system is inequitable because low-mileage drivers subsidize insurance costs for high-mileage drivers, and low-income people drive fewer miles on average.

A better approach is simple and obvious: pay-as-you-drive (PAYD) auto insurance.
With PAYD, insurance premiums would be priced per mile driven. All other risk factors will still be taken into account so that a high-risk driver would pay a higher per-mile premium than a low-risk driver. With insurance costs that vary with miles driven, people would be able to save money by reducing their driving, and this incentive would lead to fewer driving-related harms. PAYD would also be more equitable because it would eliminate the cross-subsidization of insurance costs from low-mileage to high-mileage drivers.

\cite{Parry:2005} shows that PAYD is slightly more efficient than a simple tax on vehicle miles traveled (VMT) for a given fuel reduction and even performs reasonably  well relative to a fully optimized VMT tax. Although PAYD insurance has long been advocated by transportation planners, little attention has been given to the precise design of a distance-based pricing system.

Remain of this paper is as follows: In Section 2, we present the discounted collective risk models and some useful Theorems on it. In Section 3, we enter the mileage to the discounted collective risk model to consider PAYD. Finally, some conclusions are presented in Section 4.

\section{Discounted Collective Risk Models}

Although the collective risk model seems to have many advantages, one of its drawbacks is that it overlooks the arrival time of claims and the effect of the interest rate. In property liability insurance contracts, there is always a time lag between the premium payment and claims arrival time. During this time lags, the insurer earns investment income on the unexpanded component of the premium, which is not involved in the collective risk model. So the insured is eligible to have some of this investment profit during the policy coverage period.

\begin{definition}
A discounted collective risk model (DCRM) in a specific period of time $(0; t]$, represents the total loss, $Z_t$, as the sum of a random number claims, $N(t)$, of individual present value payment amounts $\left( X_1,X_2,...,X_{N(t)}\right)$ respect to arrival times $\left( W_1,W_2,...,W_{N(t)}\right)$ and constant force of interest $\delta$ as follows:

	\begin{equation}
	\label{Dis_Coll}
	Z_t=\sum_{i=1}^{N(t)}X_i\,e^{-\delta W_i}
	\end{equation}
	Where:
\begin{itemize}
	\item 	Individual claims $X_i$ are independent and identically distributed,
	\item $N(t)$ and $X_i$ are independent, and
	\item $Z_t=0$ when $N(t)=0$.
\end{itemize}
\end{definition}	

The DCRM defined as above has several interesting and useful properties. At first, it incorporates investment income into the pricing model. Moreover, it provides a better model for property and liability insurance in which the interval between premium payments and claim payments is a significant factor. Therefore, insurers can present long term insurance products in the property and liability insurance market. One important quantity is the expected value of $Z_t$, which can be interpreted as the net premium amount needed to cover insurance liability on its becoming due without paying any expenses or contingent charges. We calculate this expect value in an important special case of DCRM where $N(t)$ has non-homogeneous Poisson distribution.

Using the martingale approach, many interesting results can be obtained; refer to \cite{Gerber:1997} for a thorough discussion. In the following theorem, we use a similar technique to find the moment generating a function of $Z_t$.

\begin{theorem}
Let $M_{Z_t}(u)$ denote the moment generating function (m.g.f) of $Z_t$ defined by relation \ref{Dis_Coll} and let $N(t)\sim Possion (\lambda(t))$, then 
\begin{equation}
\label{Moment_generatin}
M_{Z_t}(u)=exp\left\lbrace -\int_{0}^{t} \lambda(s)\left( 1-M_X\left( u \,e^{-\delta s}\right) \right)  \right\rbrace, 
\end{equation}	
where $M_X(.)$ is m.g.f of $X$.
\end{theorem}	

\noindent \textit{Proof.} Consider process $\left\lbrace M_t \right\rbrace_{t>0}=\left\lbrace \frac{e^{uZ_t}}{g(t,u)}\right\rbrace_{t>0} $ where $g(t,u)$ is a function to be determined later and satisfies in the initial condition $g(0,u)=1$. We first seek a value of $g(t,u)$ such that $\left\lbrace M_t \right\rbrace_{t>0}$ is a martingale. To do this, we note that based on the properties of martingale, $M_t$ must satisfy in the following relation (for all $h > 0$): 
\[ E\left[ \frac{M_{t+h}} {M_t}| M_t=m_t \right]=1.\]
In this case, we have:
\[ E\left[ e^{u\sum_{i=N(t)+1}^{N(t+h)} X_i e^{-\delta W_i}  }               \right]   =\frac{g(t+h,u)}{g(t,u)}         \]
Now by the rule of Iterated expectation, it can be shown that
\begin{equation}
\label{eq3}
\begin{array}{ccc}
E\left[ E\left[ e^{u\sum_{i=N(t)+1}^{N(t+h)} X_i e^{-\delta W_i}  } | N(t+h)-N(t)=k \right] \right]   \\=   \sum_{k=0}^{\infty} E\left[ e^{u\sum_{i=N(t)+1}^{N(t+h)} X_i e^{-\delta W_i}  } |N(t+h)-N(t)=k \right] Pr\left[ 
N(t+h)-N(t)=k\right] \\=\frac{g(t+h,u)}{g(t,u)}
\end{array}
\end{equation}
Based on the properties of the Poisson process we can rewrite \ref{eq3} as follows,
\begin{equation*}
\label{eq3}
\begin{array}{ccc}
\sum_{k=0}^{\infty} E\left[ e^{u\sum_{i=N(t)+1}^{N(t+h)} X_i e^{-\delta W_i}  }  |N(t+h)-N(t)=k\right] Pr\left[ 
N(t+h)-N(t)=k\right]   \\=   
\left(1-\lambda(t)h \right)+E\left[ e^{u X_{N(t+h)} e^{-\delta (t+h)} }|N(t+h)-N(t)=k \right] \lambda(t)h+o(h)\\=
\left( 1-\lambda (t)h\right) + M_X \left( u e^{-\delta (t+h)}\right)\lambda(t)h+o(h)     = \frac{g(t+h,u)}{g(t,u)}
\end{array}
\end{equation*}
 Where $o(h)$ is a generic function that goes to zero faster than $h$ when $h$ goes to zero. By a few simplifications we have:
 
 \begin{equation}
\label{eq4}
-\lambda(t) \left( 1- M_X \left( u e^{-\delta (t+h)}\right)\right) +\frac{o(h)}{h}=\frac{g(t+h,u)-g(t,u)}{h.g(t,u)}
 \end{equation}

Taking limits as $h \rightarrow 0$ in the above relation, we have:

\begin{equation}
\label{eq5}
 \frac{d}{dt}ln g(t,u)=\left( 1- M_X \left( u e^{-\delta t}\right)\right) 
\end{equation}
Now it is sufficient to show that $M_{Z_t}=g(t,u)=g(t,u)$. It follows from the initial condition $g(0,u)=1$ that $M_0=1$. Moreover based on the properties of martingale, we have $E(M_t) = E (M_0) = 1$, which completes the proof.

\begin{corollary}
Suppose in the $DCRM (1)$, $N (t) \sim Poisson (\lambda t)$, then
\begin{equation}
E\left[Z_t \right]=\frac{\mu_1 \lambda}{\delta}\left( 1-e^{-\delta t}\right),  
\end{equation}
where $\mu_1=E(X)$.
\end{corollary}

\begin{corollary}
	Process $\left\lbrace A_t \right\rbrace_{t>0}=\left\lbrace Z_t-
	\frac{\mu_1 \lambda}{\delta}\left( 1-e^{-\delta t}\right)
	\right\rbrace_{t>0}  $ is a martingale.

\end{corollary}

Let us now consider the discrepancy between the obtained premiums based on the
collective risk model, and by the DCRM equation (6). Recall that if $S_t$ shows the collective risk model, then $S_t=\sum{i=1}{N(t)} X_i$ In fact $S_t$ is a special case of relation (6) when the $\delta \rightarrow 0$. It is easy to see that, 
\[ \lim\limits_{\delta \rightarrow 0}E\left[ Z_t \right]=E\left[ S_t\right]    .\]

Another special case is when $t \rightarrow \infty$. In this case:
\begin{equation}
\label{eq7}
\lim\limits_{t \rightarrow \infty} E\left[Z_t \right]=\lim\limits_{t \rightarrow \infty} \frac{\mu_1 \lambda}{\delta}\left( 1-e^{-\delta t}\right),  
\end{equation}

which can be interpreted as a single net premium for a perpetuity that continuously pays $\mu_1 \lambda$.

Moreover, note that if $\delta \rightarrow \infty$, then $E [Z_t] \rightarrow 0$, and when $t \rightarrow  0$, then $E [Z_t] \rightarrow 0$, which are reasonable results.

\begin{corollary}
Consider the DCRM described in \refeq{Dis_Coll}, if $N(t) \sim Poisson (\lambda t)$, then

\begin{equation}
\label{eq8}
Var\left[Z_t \right]=\frac{\mu_2 \lambda}{2\delta}\left( 1-e^{-2\delta t}\right)  
\end{equation}	
where $\mu_2=E\left( X^2\right) $.
\end{corollary}

\begin{corollary}
The process $\left\lbrace B_t  \right\rbrace_{t>0}=\left\lbrace 
\left(Z_t-\frac{\mu_1 \lambda}{\delta} \left( 1-e^{-\delta t}\right)  \right)^2 -\frac{\mu_2 \lambda}{2\delta} \left(  1- e^{-2\delta t}\right)  
\right\rbrace_{t>0}  $ is a martingale.
\end{corollary}

\begin{example}
In the discounted collective risk, let claim sizes are exponentially distributed
with mean then the m.g.f is given by:
\begin{equation}
\label{eq9}
M_{Z_t}(u)=E\left[ e^{u Z_t}\right] = \left( \frac{1-\beta u e^{-\delta t} }{1- \beta u} \right)^{\frac{\lambda}{\delta}} .
\end{equation}
It follows from this example that:
\begin{equation}
\label{eq10}
\lim\limits_{\delta \rightarrow 0} M_{Z_t}(u)=exp\left( \frac{\lambda t u \beta}{1-u\beta}\right) 
\end{equation}
Moreover by limiting when t tend to innity we have:
\begin{equation}
\label{eq11}
\lim\limits_{\delta \rightarrow 0} M_{Z_t}(u)=\left(1-u\beta \right)^{-\frac{\lambda}{\delta}} 
\end{equation}
which are coincide to the results that Gerber (1979) has obtained.
\end{example}

\section{Modeling PAYD by means of DCRM}
In PAYD Insurance, we have lots of information with the help of the GPS system, and our goal is to set a premium based on the distance that a person travels during a year. By presenting this new product, insurers are facing a new source of risk, which is a random premium. We do not have an exact amount of kilometers that the driver will drive. Let $d(t)$ is mileage to time t. We would like to define a DCRM that consider $d(t)$. To do this, there is two different approaches to consider $d(t)$ in the models. 

The first approach is using double subordinated model defined by \cite{Sato:1999,Shirvani2019a}. Subordination is an often used stochastic process in modeling asset prices. Applications of subordination model and L\'{e}vy processes arise in science and engineering, e.g., quantum mechanics, insurance, economics, finance, biomathematics, etc.\footnote{See \cite{Michna:2010},  \cite{Sims:2012},\cite{Lefevre:2013}, \cite{Morales:2007}, \cite{Levajkovic:2016}, \cite{Shirvani:2019}, and \cite{ShirvaniEquity:2019}. }  \cite{Shirvani2019a} introduced the theory of multiple subordinated model to modeling the tail behavior of stock market returns.\footnote{See also \cite{Shirvani2019c}.}  

To apply the double subordinator models for modeling the the DCRM, Let $X_i$ and $d(t)=U(t)$, be  L\'{e}vy subordinators.\footnote{A L\textrm{\'{e}}vy subordinator is a L\textrm{\'{e}}vy process with an increasing sample path \citep[see][]{Sato:1999}.} Then, the  double subordinator $V(t)=X_i\left( U(t)\right)$ represent the individual claims when the subordinator $d(t)=U(t)$ is the  miles mileage to time t. Therefore, the DCRM model is 

	\begin{equation}
	\label{Dis_Coll}
	Z_t=\sum_{i=1}^{N(t)}X_i\left( U(t)\right) \,e^{-\delta W_i}. 
	\end{equation}
However, this model for DRCM is a new method, which is beyond the scope of this paper.

The second approach for modeling $d(t)$ is to use the Double Stochastic Poisson Process (DSPP). Since the goal of this paper is using DSPP processes, let us give a brief definition of DSPP. We notice that many alternative definitions of a DSPP can be given \citep[see][]{Grandell:1976,Bremaud:1981}.

\begin{definition}
A DSPP $\left\lbrace N(t): t > t_0 \right\rbrace $  with intensity stochastic process $\left\lbrace  \lambda\left( t, d(t)\right)  : t >t_0 \right\rbrace$  is defined as a conditioned Poisson process which intensity is the process $\left\lbrace \lambda\left( t, d(t))\right) :
t > t_0\right\rbrace$  given the information process $\left\lbrace d(t) : t > t_0\right\rbrace$ .
\end{definition}

The DSPP, or Cox process, provides flexibility by letting the intensity not only depend on time but also by allowing it to be a stochastic process. Therefore, the doubly stochastic Poisson process can be viewed as a two-step randomization procedure. An intensity process $\left\lbrace  \lambda\left( t, d(t)\right)  : t >t_0 \right\rbrace$ is used to generate another process $\left\lbrace N(t): t > t_0 \right\rbrace $ by acting as its intensity. If  $\left\lbrace  \lambda\left( t, d(t)\right)  : t >t_0 \right\rbrace$ is deterministic, then $\left\lbrace N(t): t > t_0 \right\rbrace $ is a nonhomogeneous Poisson process. If $\left\lbrace  \lambda\left( t, d(t)\right)  : t >t_0 \right\rbrace=\lambda$  for some positive random variable $\lambda$, then  $\left\lbrace N(t): t > t_0 \right\rbrace $ is a mixed Poisson process.

\begin{theorem}
Let $M_{Z_t} (u)$ denote the m.g.f of $Z_t$ defined by relation \eqref{Dis_Coll}  and let $N (t)$ is a DSPP with intensity process $\left( \lambda(t, d(t))\right) $, then
\begin{equation}
\label{eq12}
M_{Z_t}(u)=E\left[ exp \left\lbrace - \int_{0}^{t} \lambda\left( s,d(s)\right) \left( 1-M_x\left( ue^{-\delta s}\right) \right) dt\right\rbrace \right] 
\end{equation} 
where $M_X (.)$ is m.g.f of $X$.
\end{theorem}
\textit{Proof.} Conditioning on $\left( \lambda(t, d(t))\right) $ and using Theorem (1) results can be obtained. 

\begin{corollary}
Under conditions of the Theorem (2) we have
\begin{equation}
E\left[ Z_t\right] = \mu_1 E\left[ \int_{0}^{t} \lambda \left(s,d(s) \right) e^{-\delta s} ds \right] .
\end{equation}
\end{corollary}

\section{Conclusion}
With PAYD, insurance premiums would be priced per mile driven. All other risk factors will still be taken into account so that a high-risk driver would pay a higher per-mile premium than a low-risk driver. With insurance costs that vary with miles driven, people would be able to save money by reducing their driving, and this incentive would lead to fewer driving-related harms. PAYD would also be more equitable because it would eliminate the cross-subsidization of insurance costs from low-mileage to high-mileage drivers.

As we said, the DSPP provides flexibility by letting the intensity not only depend on time but also by allowing it to be a stochastic process. Therefore, the doubly stochastic Poisson process can be viewed as a two-step randomization procedure. We show that it is possible to model PAYD by using DSPP.

\normalem

\end{spacing}

\begin{thebibliography}{}
	
	\bibitem[\protect\citeauthoryear{Bordoff and Noel}{Bordoff and
		Noel}{2008}]{Bordoff:2008}
	Bordoff, J. and P.~Noel (2008).
	\newblock {\em Pay-As-You-Drive Auto Insurance: A Simple Way to Reduce
		Driving-Related Harms and Increase Equity}.
	\newblock The Brooking Institution.
	
	\bibitem[\protect\citeauthoryear{Bremaud}{Bremaud}{1981}]{Bremaud:1981}
	Bremaud, P. (1981).
	\newblock {\em Point Processes and Queues: Martingale Dynamics}.
	\newblock New York: Springer Verlag.
	
	\bibitem[\protect\citeauthoryear{Denuit and Dhaene}{Denuit and
		Dhaene}{2001}]{Denuit:2001}
	Denuit, M. and J.~Dhaene (2001).
	\newblock Bonus-malus scales using exponential loss functions.
	\newblock {\em Bl{\"a}tter der DGVFM\/}~{\em 25}, 13--27.
	
	\bibitem[\protect\citeauthoryear{Frangos and Vrontos}{Frangos and
		Vrontos}{2001}]{Frangos:2001}
	Frangos, N. and S.~Vrontos (2001).
	\newblock Design of optimal bonus malus systems with a frequency and severity
	component on an individual base in automobile insurance.
	\newblock {\em Astin Bulletin\/}~{\em 31\/}(1), 1--22.
	
	\bibitem[\protect\citeauthoryear{Gerber and Shiu}{Gerber and
		Shiu}{1997}]{Gerber:1997}
	Gerber, H. and S.~Shiu (1997).
	\newblock The joint distribution of the time of ruin, the surplus immediately
	before ruin, and the deficit at ruin.
	\newblock {\em Insurance: Mathematics and Economics\/}~{\em 21\/}(2), 129--137.
	
	\bibitem[\protect\citeauthoryear{Grandell}{Grandell}{1976}]{Grandell:1976}
	Grandell, P. (1976).
	\newblock {\em Doubly Stochastic Poisson Processes}.
	\newblock Berlin: Springer Verlag.
	
	\bibitem[\protect\citeauthoryear{Lef\'{e}vre and Picard}{Lef\'{e}vre and
		Picard}{2013}]{Lefevre:2013}
	Lef\'{e}vre, C. and P.~Picard (2013).
	\newblock Ruin time and severity for a l\'{e}vy subordinator claim process: A
	simple approach.
	\newblock {\em Journal of Risks\/}~{\em 1}, 192–212.
	
	\bibitem[\protect\citeauthoryear{Lemaire}{Lemaire}{1985}]{Lemaire:1985}
	Lemaire, J. (1985).
	\newblock {\em Automobile Insurance: Actuarial Models}.
	\newblock Netherlands: Kluwer Nijholff.
	
	\bibitem[\protect\citeauthoryear{Lemaire}{Lemaire}{1995}]{Lemaire:1995}
	Lemaire, J. (1995).
	\newblock {\em Bonus Malus Systems in Automobile Insurance}.
	\newblock Boston: Kluwer Academic Publisher.
	
	\bibitem[\protect\citeauthoryear{Levajkovi\'{c}, Mena, and
		Zarfl}{Levajkovi\'{c} et~al.}{2016}]{Levajkovic:2016}
	Levajkovi\'{c}, T., H.~Mena, and M.~Zarfl (2016).
	\newblock L\'{e}vy processes, subordinators and crime modeling.
	\newblock {\em Novi Sad J. Maths\/}~{\em 46}, 65–86.
	
	\bibitem[\protect\citeauthoryear{Litman}{Litman}{2009}]{Litman:2009}
	Litman, T. (2009).
	\newblock Pay-as-you-drive pricing for insurance affordability.
	\newblock {\em Victoria Transport Policy Institute\/}, 1--19.
	
	\bibitem[\protect\citeauthoryear{Mahmoudvand and Hassani}{Mahmoudvand and
		Hassani}{2009}]{Mahmoudvand:2009}
	Mahmoudvand, R. and H.~Hassani (2009).
	\newblock Generalized bonus-malus systems with a infrequence and severity
	component on an individual basis in automobile insurance.
	\newblock {\em ASTIN Bulletin\/}~{\em 39}, 307--315.
	
	\bibitem[\protect\citeauthoryear{Michna}{Michna}{2010}]{Michna:2010}
	Michna, Z. (2010).
	\newblock Ruin probability on a finite time horizon.
	\newblock {\em Math. Econ.\/}~{\em 6}, 65–74.
	
	\bibitem[\protect\citeauthoryear{Morales}{Morales}{2007}]{Morales:2007}
	Morales, M. (2007).
	\newblock On the expected discounted penalty function for a perturbed risk
	process driven by a subordinator.
	\newblock {\em Insur. Math. Econ.\/}~{\em 40}, 293–301.
	
	\bibitem[\protect\citeauthoryear{Parry}{Parry}{2005}]{Parry:2005}
	Parry, T. (2005).
	\newblock {\em Is Pay-As-You-Drive Insurance a Better Way to Reduce Gasoline
		than Gasoline Taxes?}
	\newblock Resources for the Future.
	
	\bibitem[\protect\citeauthoryear{Sato}{Sato}{1999}]{Sato:1999}
	Sato, K.-I. (1999).
	\newblock {\em L\`{e}vy Processes and Infinitely Divisible Distributions}.
	\newblock Cambridge: Cambridge studies in advanced mathematics.
	
	\bibitem[\protect\citeauthoryear{Shirvani, Hu, Rachev, and Fabozzi}{Shirvani
		et~al.}{2019}]{Shirvani2019c}
	Shirvani, A., Y.~Hu, S.~Rachev, and F.~Fabozzi (2019).
	\newblock Mixed levy subordinated market model and implied probability
	weighting function.
	\newblock {\em arXiv:1910.05902\/}.
	
	\bibitem[\protect\citeauthoryear{Shirvani, Rachev, and Fabozzi}{Shirvani
		et~al.}{2019}]{Shirvani2019a}
	Shirvani, A., S.~Rachev, and F.~Fabozzi (2019).
	\newblock Multiple subordinated modeling of asset returns.
	\newblock {\em arXiv:1907.12600\/}.
	
	\bibitem[\protect\citeauthoryear{Shirvani, Stoyanov, Fabozzi, and
		Rachev}{Shirvani et~al.}{2019}]{ShirvaniEquity:2019}
	Shirvani, A., S.~Stoyanov, F.~Fabozzi, and S.~Rachev (2019).
	\newblock Equity premium puzzle or faulty economic modelling?
	\newblock {\em arXiv:1909.13019\/}.
	
	\bibitem[\protect\citeauthoryear{Shirvani and Volchenkov}{Shirvani and
		Volchenkov}{2019}]{Shirvani:2019}
	Shirvani, A. and D.~Volchenkov (2019).
	\newblock A regulated market under sanctions: On tail dependence between oil,
	gold, and tehran stock exchange index.
	\newblock {\em Journal of Vibration Testing and System Dynamics\/}~{\em
		3\/}(3), 297--311.
	
	\bibitem[\protect\citeauthoryear{Sims, Humphries, Bradford, and Bruce}{Sims
		et~al.}{2012}]{Sims:2012}
	Sims, D., N.~Humphries, R.~Bradford, and B.~Bruce (2012).
	\newblock L\'{e}vy flight and brownian search patterns of a free-ranging
	predator reflect different prey field.
	\newblock {\em J Anim Ecol\/}~{\em 81\/}(2), 432–442.
	
	\bibitem[\protect\citeauthoryear{Taylor}{Taylor}{1997}]{Taylor:1997}
	Taylor, G. (1997).
	\newblock Setting a bonus-malus scale in the presence of other rating factors.
	\newblock {\em ASTIN Bulletin\/}~{\em 27}, 319--327.
	
\end{thebibliography}
\end{document}